\documentclass[iop]{emulateapj}

\newcommand{\hi}{\protect\ion{H}{1}}
\newcommand{\msun}{$M_{\odot}$}

\newcommand{\kms}{~km\,s$^{-1}$}

\begin{document}

\title{Evolution in the \hi\ Gas Content of Galaxy Groups: \\
Pre-processing and Mass Assembly in the Current Epoch}

\author{Kelley M. Hess\altaffilmark{1,2}} \affil{Astrophysics, Cosmology and Gravity Centre (ACGC), \\ Department of Astronomy, University of Cape Town, Rondebosch 7701, South Africa;\\
  hess@ast.uct.ac.za}
\and
\author{Eric M. Wilcots\altaffilmark{2}} \affil{Department of Astronomy,
  University of Wisconsin-Madison, Madison, WI 53706, USA;\\
  ewilcots@astro.wisc.edu}

\begin{abstract}
We present an analysis of the neutral hydrogen (\hi) content and distribution of galaxies in groups as a function of their parent dark matter halo mass.  The Arecibo Legacy Fast ALFA survey $\alpha.40$ data release allows us, for the first time, to study the \hi\ properties of over 740 galaxy groups in the volume of sky common to the Sloan Digital Sky Survey and ALFALFA surveys.  We assigned ALFALFA \hi\ detections a group membership based on an existing magnitude/volume-limited SDSS DR7 group/cluster catalog.  Additionally, we assigned group ``proximity'' membership to \hi\ detected objects whose optical counterpart falls below the limiting optical magnitude--thereby not contributing substantially to the estimate of the group stellar mass, but significantly to the total group \hi\ mass.  We find that only 25\% of the \hi\ detected galaxies reside in groups or clusters, in contrast to approximately half of all optically detected galaxies.  Further, we plot the relative positions of optical and \hi\ detections in groups as a function of parent dark matter halo mass to reveal strong evidence that \hi\ is being processed in galaxies as a result of the group environment: as optical membership increases, groups become increasingly deficient of \hi\ rich galaxies at their center and the \hi\ distribution of galaxies in the most massive groups starts to resemble the distribution observed in comparatively more extreme cluster environments.  We find that the lowest \hi\ mass objects lose their gas first as they are processed in the group environment, and it is evident that the infall of gas rich objects is important to the continuing growth of large scale structure at the present epoch, replenishing the neutral gas supply of groups.  Finally, we compare our results to those of cosmological simulations and find that current models cannot simultaneously predict the \hi\ selected halo occupation distribution for both low and high mass halos.
\end{abstract}

\section{Introduction}

Observationally, galaxies are the reservoirs of luminous baryonic matter that allow us to trace the large scale structure of the Universe.  Galaxies exhibit a wide range of properties, and it is clear that environment, as measured by the density of galaxies per Mpc$^3$, is linked to trends in galaxy morphology, color, age, and star formation rates.  It has long been known that the cluster population is fundamentally different from galaxies that reside in the field \citep{Hubble31}, and the morphology-density relation quantified how the fraction of S0 and early-type galaxies increases
with increasing density environment \citep{Dressler80}.  Even within a given morphological type the stellar content of galaxies exhibits environmental trends.  For example, cluster early-type galaxies appear to be older than their counterparts in lower density environments \citep{Bower90,Rose94}, and
\citet{Balogh97} conclude that the fraction of star forming galaxies is suppressed in the cluster environment relative to the field (see also \citealt{Lewis02,Gomez03}).

Despite the correlation between galaxy properties and their environment, it is still unclear to what degree the environment drives evolution (nurture) versus being a passive tracer of innate galaxy properties (nature).  Detailed work at optical wavelengths has depicted the variation of the stellar content with environment which suggests galaxies undergo transformation on the outskirts of clusters and in intermediate density environments (e.g.~\citealt{Homeier06,Porter07,Lane07}).  The galaxy transformation is likely linked to the removal or exhaustion of a galaxy's gas supply, limiting its ability to form new stars, in addition to its morphological transformation.  \citet{Goto04} find evidence that morphological transformation from late- to early-type galaxies works on longer timescales than color evolution from the blue cloud to the red sequence.  

In order to gain a full understanding of galaxy evolution, we require a complete census of the gas in galaxies in addition to the stars.  As the component of galaxies most easily disturbed by changes in environment, neutral hydrogen gas (\hi) is a tracer for recent and ongoing interactions between a galaxy and its surroundings, as well as a measure of its star formation potential.  Until recently, the impact of environment on galaxy evolution from an \hi\ perspective has been dominated by pointed observations of optically bright, late-type galaxies with respect to extreme environments of clusters or the field (e.g.~\citealt{Gavazzi06,Chung09}), although a growing number of studies also examine \hi\ in well defined intermediate group environments (e.g.~\citealt{Pisano07,Kern08,Freeland09,Kilborn09,Rasmussen12,Mihos12}). These studies have shown that galaxies in the dense cluster environment tend to be \hi\ deficient (e.g.~\citealt{Sullivan81,Haynes84,BravoAlfaro00,Solanes01}), and that the distribution of \hi\ mass in galaxies, formulated in the \hi\ mass function, likely evolves with environment \citep{Rosenberg02,Springob05}.

A number of physical processes may be responsible for driving the gaseous evolution of galaxies which is then reflected in their stellar properties.  These mechanisms fall into either the category of gravitational or hydrodynamical interactions, and different mechanisms are more efficient in different environments.  In clusters, the environment is conducive to galaxy harassment from fast, minor gravitational encounters between galaxies \citep{Moore98}, tidal interaction between galaxies and the intracluster potential \citep{Moore99}, starvation as the environment strips off
a galaxy's supply of hot gas preventing it from reacquiring gas with which to make stars \citep{Larson80}, and ram pressure stripping of the interstellar medium of a galaxy as it moves through the hot intracluster medium (\citealt{Gunn72}; \citealt{Hester06} and references therein). Dramatic observations of ram pressure stripping demonstrate how galaxies falling into the cluster experience gas removal as far out as $1-2$ virial radii (e.g.~\citealt{Kenney04,Crowl05,Chung09,Abramson11,Bahe13}) eventually leading to the shut down of star formation.  It is well established that galaxies in the Coma Cluster are \hi\ deficient out to 1 virial radius \citep{Sullivan81,BravoAlfaro00,Gavazzi06}.  Beyond this, it is claimed that galaxies in the supercluster filament are indistinguishable from field galaxies \citep{Chincarini83,Solanes01,Gavazzi06}.  The combination of optical and \hi\ observations suggests that galaxies may undergo rapid transition at the boundary of clusters. 

However, there is evidence that galaxies must have undergone some amount of ``pre-processing'' before falling into clusters (e.g.~\citealt{Balogh97,Zabludoff98,Mahajan13}).  Optical and infrared all-sky redshift surveys have revealed that half to two-thirds of all galaxies reside in groups
\citep{Huchra82,Eke04,Crook07,Berlind06}, collections of a few to a hundred gravitationally bound members, and it follows that this environment is likely where galaxies undergo the majority of their evolution.  Unfortunately, groups are difficult to identify and study: they do not have the same dramatic spatial concentration of clusters, and with few members, their dark matter halos are poorly populated requiring large samples of groups in order to draw statistically significant conclusions.  Construction of mock group catalogs demonstrates the impossibility of recovering all group parameters simultaneously (velocity dispersion, projected size, multiplicity function, and occupation distribution; \citealt{Berlind06}).  Finally, in the poorest groups, it can be difficult to determine whether galaxies are gravitationally bound, or a chance superposition of galaxies with similar apparent velocities.

The physical processes which dominate gas evolution, trigger star formation, or feed active galactic nuclei activity are different in groups compared to clusters because the physical conditions are different: velocity dispersion, density of members, thermodynamics of the intergalactic medium, etc.  The low velocity dispersion, of order the rotational speed of a galaxy, combined with the close proximity of group members, means gravitational encounters and major mergers occur more frequently in groups compared to clusters.  In X-ray bright groups strangulation and evaporation are likely important in removing the hot gaseous halo of galaxies which would otherwise provide a reservoir of cooling gas to feed star formation in the disk.  Evidence for this is found in groups with X-ray emission which tend to be more \hi\ deficient than those without \citep{Kilborn09,Rasmussen12,Desjardins13}.  Some of this cold gas loss may also be attributed to viscous stripping \citep{Nulsen82} and/or ram pressure stripping despite the presumed lower intargroup medium temperatures and densities, but so far observations have been inconclusive.  

In any case, population studies derived from small samples of groups consistently show that the group \hi\ mass function (HIMF) is flatter than the global HIMF, revealing either a paucity of low gas-mass galaxies or simply hinting that denser environments are harsh on low gas-mass galaxies
\citep{Verheijen00,deBlok02,Kovac05,Freeland09,Kilborn09,Pisano11,Hess11}.  While \citet{Freeland09} found that groups with a mix of morphological types exhibit more signs of recent tidal interactions than groups dominated by one morphological type. Thus, collectively and over varying timescales, gravitational and to a lesser extent, hydrodynamical interactions can also transform galaxies from blue, star-forming, gas rich objects to red, passively evolving, gas poor objects in groups. Identifying how and where this transition occurs in groups has proven difficult.

Gas removal is important for shutting off star formation at late times and in dense environments, but the importance of continued infall of gas rich galaxies is an unanswered question \citep{BravoAlfaro00}.  Cosmological simulations successfully reproduce the clustering of dark matter halos at the present epoch (e.g.~\citealt{Springel05}), but populating the halos with a realistic distribution of galaxies is more difficult.  Part of the answer may lie in accurately modeling the merger history of galaxies within halos as well as the merger of group sized halos to form larger structures.  On galaxy scales,
semi-analytical models suggest that mergers are required to explain the observed luminosity, morphology, and color distribution of galaxies \citep{Kauffmann93}.  Meanwhile, models disagree on the contribution of mergers to growth on large scales.  As little as 12\% or as much as 50\% of cluster {\it stellar} mass may have assembled through the accretion of groups \citep{Berrier09,McGee09}.  Unfortunately, gas treatments of galaxy evolution in these cosmological simulations are plagued by the complications of star formation and AGN feedback, but efforts are being made (e.g.~\citealt{Kim11}).  Observations remain the best insight to the gaseous evolution of galaxies.

Until recently, we lacked a complementary wide-field redshift survey of the cold, neutral \hi\ gas content of galaxies with sufficient resolution and sensitivity to examine the detailed gas distribution of galaxies without dedicated interferometric follow-up.  The Arecibo Legacy Fast ALFA (Arecibo L-band Feed Array) survey is a blind, drift scan survey of over 7000 deg$^2$ of the sky accessible to the Arecibo Observatory designed to measure the neutral hydrogen (\hi) content of the largest over and under dense environments in the local Universe \citep{Giovanelli05} whose spatial coverage is well-matched to the optical Sloan Digital Sky Survey.  ALFALFA contributions include thousands of new \hi\ detections out to $cz_{\odot}=18,000$ km s$^{-1}$, with significant mass sensitivity and resolution to be valuable as a tool for multi-wavelength data mining \citep{Haynes11}. The combination of this survey with 2MASS, SDSS photometry and spectroscopy, GALEX, and other surveys permits comprehensive multi-wavelength studies of the stellar and gas content of galaxies
across all range of environments.

In this paper we present a detailed study of the gas content of galaxy groups as a function of their parent dark matter halo mass.  We demonstrate that there is already substantial evolution in the \hi\ content of galaxies with the growth of large scale structure in intermediate mass halos.  Further, we spatially resolve the \hi\ content of galaxy groups and find evidence that the physical mechanisms for removing gas which are at work on the outskirts of
clusters, may also be important in the group environment contributing to the ``pre-processing'' of galaxies.  Finally we attempt to present our results in a manner that will be easily comparable to cosmological simulations which model the cold gas content of galaxies at the present epoch.  By utilizing blind all-sky surveys we capture the stellar and gas properties across a range of galaxy populations, from previously known gas rich objects, to galaxies in transition from actively star forming to passively evolving through the full volume that overlaps between SDSS and the ALFALFA $\alpha.40$ surveys.  Ultimately, we seek to understand how much ``pre-processing'' of galaxies occurs in groups before they fall into clusters.

This paper is organized as follows: in Section \ref{data} we present the SDSS DR7 group/cluster catalog from which we define galaxy groups and we describe the \hi\ data from the ALFALFA survey; in Section \ref{results} we describe how the \hi\ distribution of galaxies varies in groups with halo occupation number; in Section \ref{discussion} we discuss the implications for galaxy evolution and galaxy transition during the build-up of hierarchical structure in the Universe; we conclude in Section \ref{conclusions}.

\section{Data}
\label{data}

In the era when all-sky and wide-field redshift surveys exist across a range of wavelengths it has become possible to construct increasingly complete group and cluster catalogs of the nearby Universe (e.g.~\citealt{Huchra82,Crook07,Berlind06}).  The most sensitive group/cluster catalogs are derived from the Sloan Digital Sky Survey (SDSS).  The optical data traces the stellar component of galaxies which effectively reveals the filamentary large scale structure \citep{Stoughton02}.  The ALFALFA survey shows that \hi\ detections trace the large scale structure network of supercluster clumps and filaments, however, in dense environments where \hi\ gas is easily removed from galaxies, simulations and observations show that stellar light better traces the mass distribution of dark matter halos (e.g.~\citealt{Kauffmann06,Haynes11,Kim11}).  As a result, we rely on the group/cluster catalog derived from SDSS Data Release 7 (DR7) $r$-band photometry and spectroscopy to provide the framework of underlying structure to which we assign group membership to ALFALFA \hi\ detections.

\subsection{SDSS groups}
\label{sdss}

The SDSS DR7 group/cluster catalogs are based on a friends-of-friends algorithm originally applied to DR4 and described in detail in \citet{Berlind06}.  The complete group/cluster and galaxy membership catalogs derived from SDSS DR7 are available online\footnote{\url{http://lss.phy.vanderbilt.edu/groups/dr7/}} \citep{Berlind09}.  Three group catalogs were created using volume-limited galaxy samples by choosing three limiting $r$-band magnitudes and calculating the redshift at which SDSS is complete for each (see Figure 1 of \citealt{Berlind06}).  We consider the ``SDSS DR7 Mr18'' sample, the faintest magnitude-limited group catalog ($M_r < -18$) which is complete to $z=0.042$.  It includes the lowest mass galaxies and as such, the group finding algorithm finds more low mass groups compared to the other group/cluster magnitude/volume-limited catalogs.  The catalog also has a redshift lower limit of $z=0.02$ to avoid problems with SDSS photometry for nearby, extended galaxies.  The limits correspond to $cz_{\odot}=12,600$\kms\ and $cz_{\odot}=6000$\kms\, respectively, which conveniently fall within the bandpass of ALFALFA and avoid the region that is significantly effected by radio frequency interference \citep{Martin10}.

The distance and velocity linking lengths for the Mr18 group catalog are based on an analysis of mock catalogs generated by populating dark matter halos in an $\Lambda$CDM cosmological $N$-body simulation.  The optimal linking length parameters are independent of the halo occupation distribution, however they have been optimized for the underlying definition of dark matter halos \citep{Berlind06}.  The linking projected distance and velocity are $(D_0, V_0)=(0.69$ Mpc, $259$ \kms). 

\subsection{ALFALFA}
\label{alfalfa}

The most recent data release for ALFALFA, the $\alpha.40$ catalog, includes 40\% of the total expected survey coverage and provides crossmatching between \hi\ sources and optical counterparts in SDSS photometric and spectroscopic surveys \citep{Haynes11}. ALFALFA detections have an approximate signal-to-noise threshold of 6.5$\sigma$, although some lower signal-to-noise detections are included in the catalog based on the additional information from SDSS
photometric and spectroscopic surveys. For a complete description of the ALFALFA survey and a discussion of crossmatching optical counterparts please see \citet{Haynes11}.

Galaxies in the volume of the Mr18 group/cluster catalog, between $6000-12,600$\kms\ are unresolved by Arecibo, but the combination of spatial and spectral resolution means few objects suffer from source confusion.  Table 2 of \citet{Haynes11} $\alpha.40$ catalog includes notes on sources for which the \hi\ parameters are uncertain because of radio frequency interference or potentially blended sources, or for uncertainties in the optical counterparts either because they are at a large distance from the \hi\ centroid or in a crowded field.  ALFALFA pointing errors are on average 18 arcseconds, and in rare occasions reach 1 arcminute, a fraction of the $3.5\times3.8$ arcminute beam.  Of the 6515 \hi\ sources in the redshift-limited volume (see Section \ref{coverage}), 324 (5.0\%) are described as possible ``blends''.  The ALFALFA team has attempted to separate these blends where possible, in which case Table 2 references another cataloged \hi\ detection and its angular distance.  In addition, 147 entries (2.3\%) are uniquely identified (not also described as ``blended'') to be in a ``crowded field'', ``interacting'', a ``pair'', or a ``group''.  These descriptions of the optical field come from examining a region of order $10\times10$ arcminutes around the \hi\ detection and they are generally more crowded than the average environment of the SDSS groups described in Section \ref{sdss}.  These \hi\ detections may be the most likely candidates to suffer from serious confusion. Due to their small numbers, we do not expect them to strongly bias our results.  As will be discussed in Section \ref{evolution}, we find that, within a given redshift shell, uncertainty associated with confusion is more common in smaller groups and removing this uncertainty would actually enhance our conclusions.

\begin{figure*}
\centering
\includegraphics[scale=0.5]{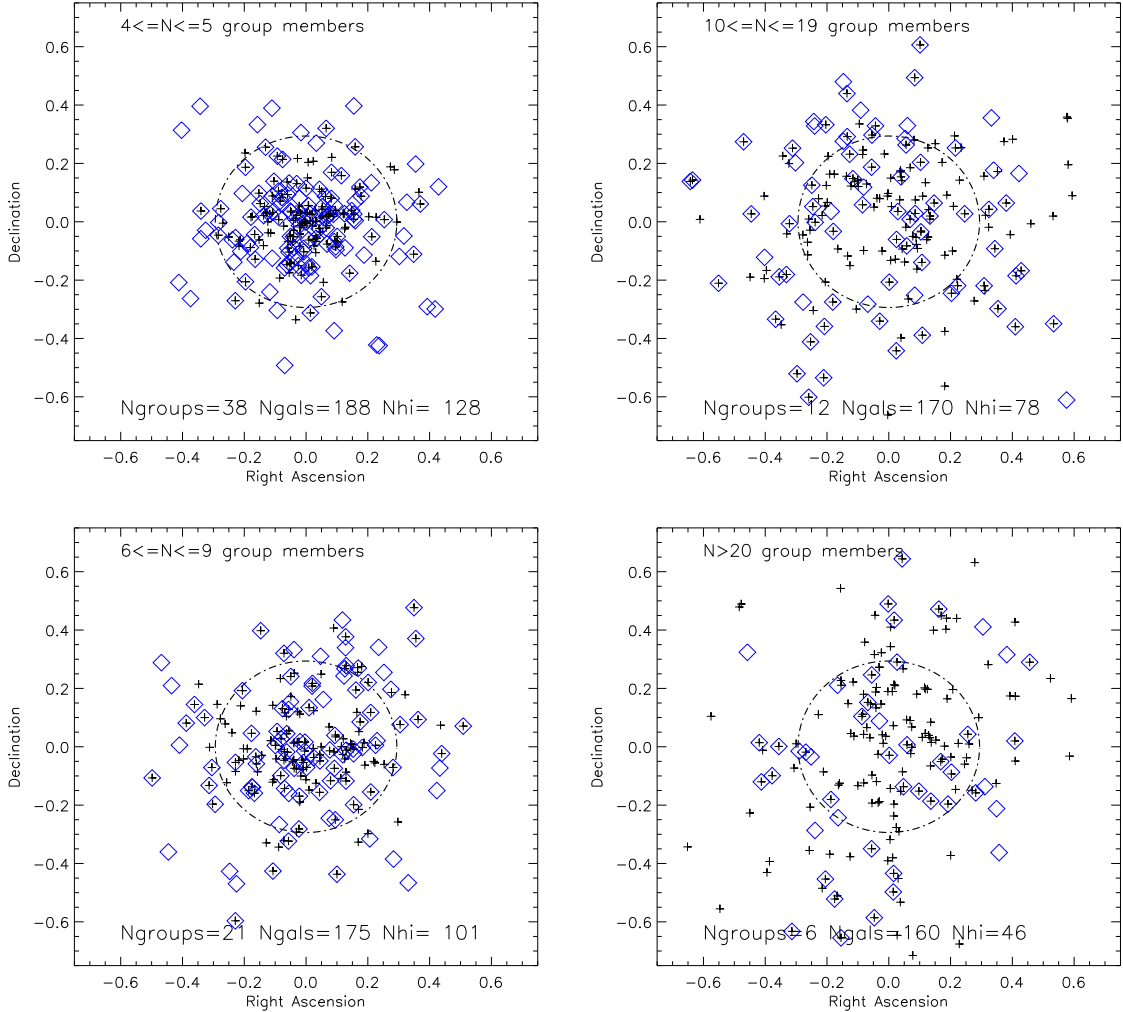}
\caption{Top: the relative positions of group member galaxies are plotted for groups in the redshift shell $cz_{\odot}=6,000-7,650$\kms. Black crosses correspond to Mr18 group members. Blue diamonds correspond to ALFALFA detected galaxies; those which do not contain black crosses are group members determined by proximity to the group center (see Section \ref{membership}). The dot-dashed circle corresponds to a circle of 1 Mpc in diameter.}
\label{pos67}
\end{figure*}

\begin{figure*}
\centering
\includegraphics[scale=0.5]{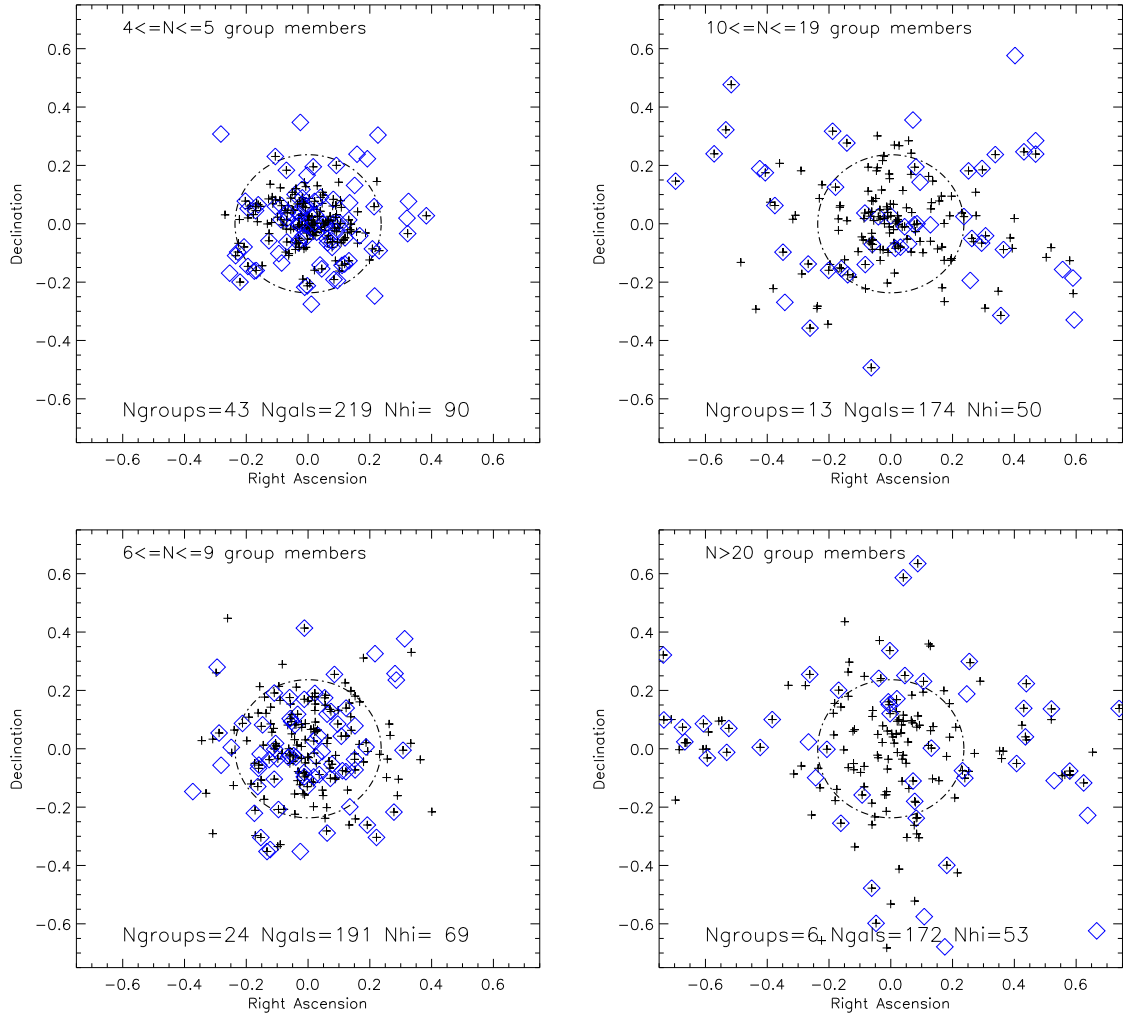}
\caption{Same as Figure \ref{pos67} but for $cz_{\odot}=7,650-9,300$\kms.}
\label{pos79}
\end{figure*}

\begin{figure*}
\centering
\includegraphics[scale=0.5]{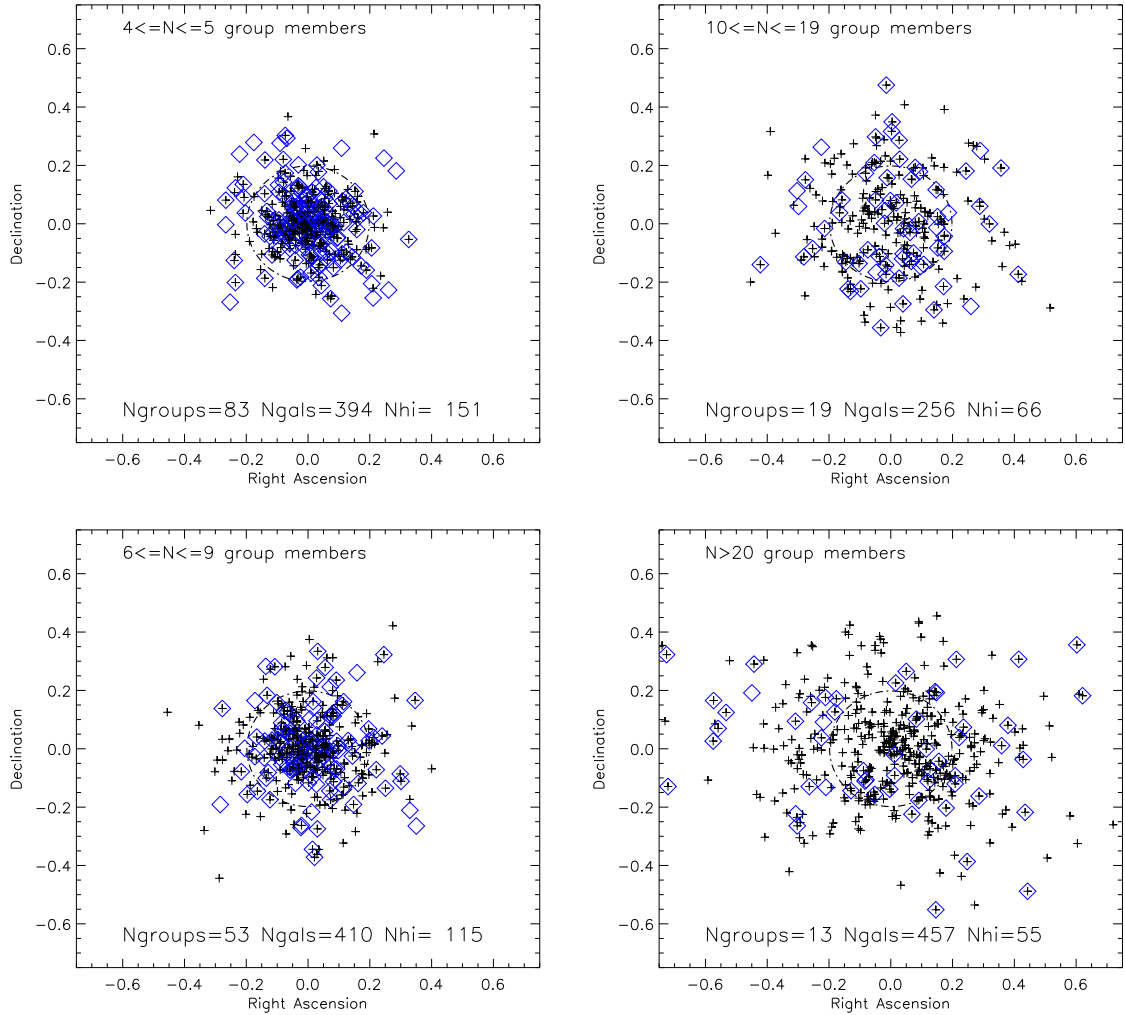}
\caption{Same as Figure \ref{pos67} but for $cz_{\odot}=9,300-10,950$\kms.}
\label{pos910}
\end{figure*}

\begin{figure*}
\centering
\includegraphics[scale=0.5]{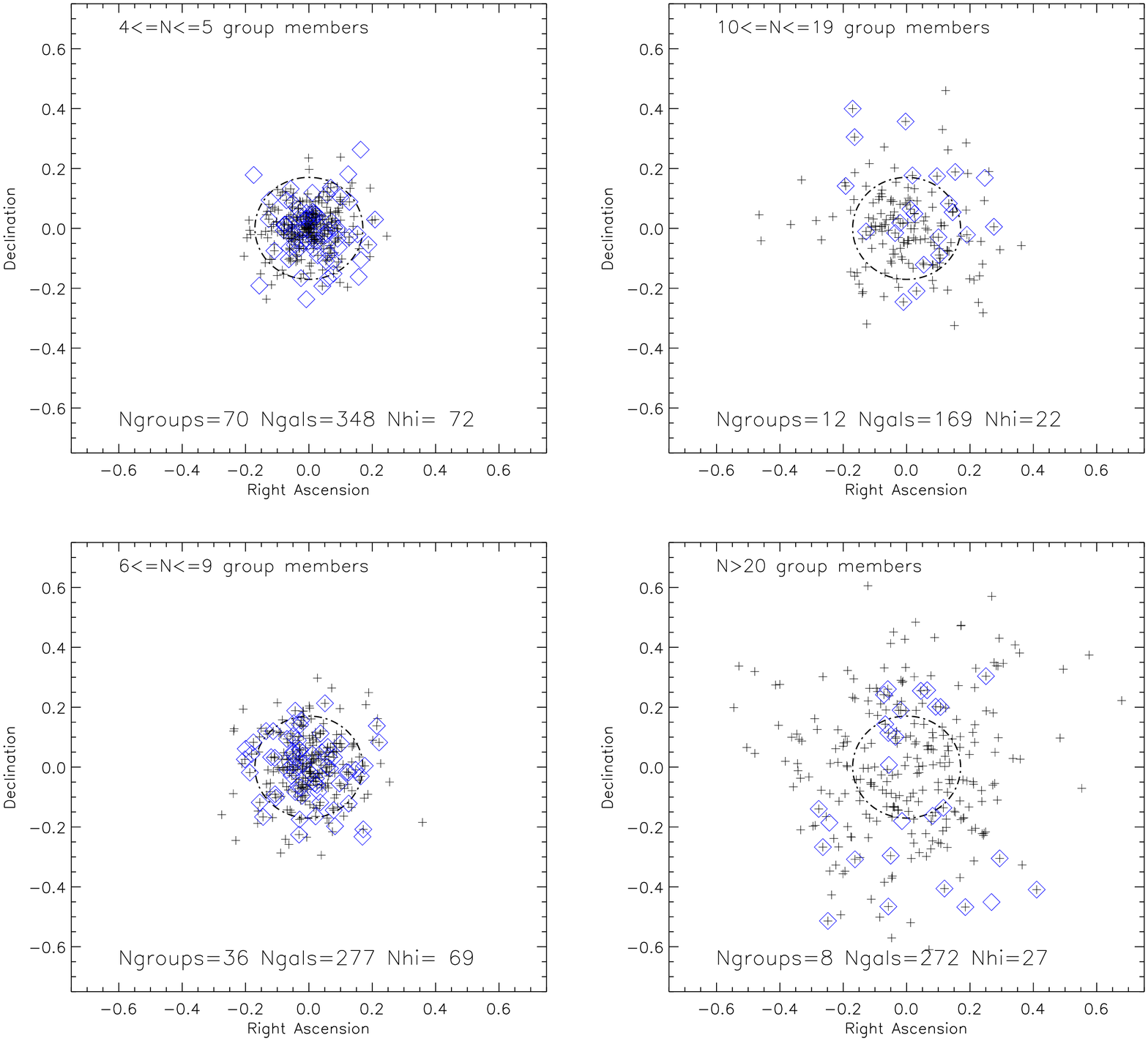}
\caption{Same as Figure \ref{pos67} but for $cz_{\odot}=10,950-12,600$\kms.}
\label{pos1012}
\end{figure*}

\subsection{Sky Coverage}
\label{coverage}

SDSS DR7 completed the photometric and spectroscopic coverage of SDSS \citep{Abazajian09}, and includes the entire Northern Galactic Cap (NGC) and a strip towards the Southern Galactic Cap (SGC).  ALFALFA is well matched to the SDSS coverage of the NGC through the full declination range accessible to Arecibo\footnote{\url{http://egg.astro.cornell.edu/alfalfa/images/alfalfa\_sdss\_coverage.jpg}}, but because of the sparse coverage by SDSS DR7 towards the SGC, we exclude that region from our analysis.  Therefore, in this paper we examine groups which fall within the ALFALFA ``Spring'' sky released thus far, covering approximately 2160 square degrees over $07^h30^m <$ R.A. $< 16^h30^m$, $+04^{\circ} <$ decl. $< +16^{\circ}$, and $+24^{\circ} <$ decl. $< +28^{\circ}$.  This overlapping volume includes 781 Mr18 group/clusters with at least three members in the redshift range between $cz_{\odot}=6000-12,600$\kms.

For simplicity in dealing with ``boundary effects'', we exclude all groups with fewer than ten members for which a single member lies outside the overlapping RA, Dec, velocity volume defined above and therefore could not have been included in the $\alpha.40$ catalog.  The sample contains two groups of 10-19 total members which have 1 or 2 optical members lying outside of the field, and four groups of greater than 20 members which have 1-7 members outside the field, but never more than 25\% of the group membership.  We choose to include these six groups in our analysis.  As will become clear, galaxies that would have been detected in \hi\ on the outskirts of these groups would only enhance our conclusions.  

Ultimately, we are left with 742 Mr18 groups of three or more members which contain 4852 optical galaxies with $M_r<-18$.  From the ALFALFA catalog there are 6515 \hi\ detected sources in the same redshift-limited volume, all of which have photometric counterparts.  Of these 925 (14\%) lack spectroscopic counterparts in DR7, but the \hi\ profiles provide a measure of their recession velocity.  The $\alpha.40$ catalog coverage towards the NGC includes the northern portion of the Coma-Abell 1367 supercluster, the largest nearby structure in the Mr18 redshift range, as well as surrounding voids, meaning the groups themselves probe all range of cosmological environments.

\subsection{HI detections in groups}
\label{membership}

Using the Mr18 catalog, we assigned \hi\ detections to group/cluster dark matter halos in a two step process.  First, we retrieved the photometric object identification (PhotoObjID) numbers for the Mr18 group members from the SDSS CasJobs website\footnote{\url{http://casjobs.sdss.org/CasJobs/}} and matched them directly to ALFALFA $\alpha.40$ sources assigned with the same photometric identification (Table 3 of \citealt{Haynes11}).  We find that 1204 ALFALFA detections correspond to Mr18 group member galaxies.

The second step uses a proximity search to identify gas rich objects in ALFALFA which either lack SDSS spectroscopic targets or whose optical counterparts are below the Mr18 magnitude limit, but which are likely group members based on their \hi\ position and systemic velocity.  The proximity search was done in a simple way by assuming an ALFALFA galaxy is a group member if it lies within a group-sized cylinder. Unfortunately, for groups with fewer than about 10 members the velocity distribution is not well sampled to assume the dispersion is representative of the group.  The problem is less severe for the projected radius.  Thus, we assume a minimum cylinder size and scale larger groups by their group properties reported in \citet{Berlind09}.  The radius of the search cylinder is chosen to either be 0.77 Mpc, or twice the mean projected radius determined from the optical membership, whichever is larger.  Similarly, the minimum velocity range is chosen to be $\pm300$\kms\, or $\pm$ the velocity dispersion of the group, whichever is larger.  Only 18 groups of the 742 have velocity dispersions larger than the minimum.  Varying the radial or velocity linking lengths by of order 15\% only produces an approximately 2\% change in group membership by adding or removing proximity members. Thus, a total of 1613 \hi\ sources are detected in 620 Mr18 groups.  

Most remarkably, only 24.8\% of ALFALFA detected \hi\ sources in the redshift-limited volume are assigned group/cluster membership.  This is in contrast to \citet{Berlind06} who applied the friends-of-friends algorithm on a similarly volume and magnitude-limited SDSS DR4 sample and report an optical group occupancy of 43.2\%, suggesting that already the group environment has modified the \hi\ content of galaxies compared to their counterparts in the field.  Previous group studies have found as much as half to two-thirds of all infrared or optically detected galaxies \citep{Huchra82,Eke04,Crook07}.

One may reasonably expect that confusion within the Arecibo beam is a greater problem in the group environment where galaxies are in close proximity.  Similarly to the investigation in Section \ref{alfalfa}, we find 147 (8.8\%) of the \hi\ group members are catalogued as ``blended'', and 46 (2.7\%) are in crowded fields or may be associated with pairs or compact groups of galaxies.  As will be shown in Section 3.2, we find that confusion is a greater issue in small groups than large ones and that the ability to correct for confusion or resolve additional sources would enhance our conclusions.

\section{Results}
\label{results}

Increasingly, the parent dark matter halo mass in which a galaxy resides is found be the most important environmental factor in determining the galaxy's evolution, after accounting for stellar mass (e.g.~\citealt{Weinmann06,Blanton07}).  Dark matter halos are easy to trace in simulation, but difficult to identify in observations.  To connect observations to simulations, we make the assumption that optical group membership is an accurate reflection of the underlying dark matter halo mass.  This is equivalent to
simulations which use a conditional luminosity function to construct the halo occupation distribution (HOD): the number of galaxies above a certain luminosity that live in a dark matter halo of a given mass. The complicated astrophysics of galaxy and structure formation guarantee that there is not a one-to-one correlation between stellar mass or group membership, and halo mass (e.g.~\citealt{Berlind03,Yang05}).  However, with a statistically significant sample of 742 groups we can finally begin to understand how gas rich galaxies are spatially distributed in groups and how the total \hi\ content varies with the average local (intragroup) environment. 

\subsection{Distribution of \hi\ in groups}

We divide the SDSS Mr18 groups into six dark matter halo mass bins based on the group/cluster catalog membership: (1) groups with three members; (2) groups with four or five optical members; (3) groups with $6\le N\le9$; (4) groups with $10\le N\le19$; and (5) groups with $N\ge20$.  Groups with 10 optical members or more are the strongest group candidates as the algorithm by \citet{Berlind06} was optimized for finding groups of that size range. Groups with six to nine members were not optimized for, but still have enough members to be satisfactory group candidates.  Groups with four or five members are questionable, but make up the low end of the group dark matter halo distribution.  Groups with three members may or may not be true groups as it is arguable as to whether they are gravitationally bound.  (6) The Coma Cluster is the only cluster in this volume.  It has 252 members which meet the optical criteria and it is discussed in section \ref{coma}.

The magnitude/volume-limited nature of the Mr18 catalog is such that groups consist of galaxies with the same optical properties throughout the entire volume from 6000 to 12,600\kms.  However, the \hi\ mass sensitivity of ALFALFA falls of with the square of the distance to the object.  To both combat the variation in \hi\ mass sensitivity through the volume and to better visualize our results, we break down our sample into four redshift shells based on the group velocity: (1) $6,000-7,650$\kms, (2) $7,650-9,300$\kms, (3) $9,300-10,950$\kms, and (4) $10,950-12,600$\kms.  In the resulting redshift shells the mass sensitivity varies by approximately $\pm$12\%, 10\%, 8\%, 7\%, respectively.

Figures \ref{pos67} - \ref{pos1012} show the spatial distribution of the optical and \hi\ detected galaxies as a function of group membership (by panel) and redshift (by figure).  To produce these figures, we shifted all group centers to the origin ($00^{h}00^{m}00^{s} +00^{\circ}00^{\prime}00^{\prime\prime}$) and plotted the relative location of their constituent galaxies. The group center is the geometric mean of group member galaxies derived by \citet{Berlind09}.  Black crosses correspond to Mr18 group members and blue diamonds correspond to ALFALFA detected galaxies.  Blue diamonds which do not contain crosses are group members determined by proximity matching.  The dot-dashed circles are equivalent to 1 Mpc diameter at the distance of the group.  

\subsubsection{The Coma Cluster}
\label{coma}
\begin{figure}
\centering
\includegraphics[scale=0.5]{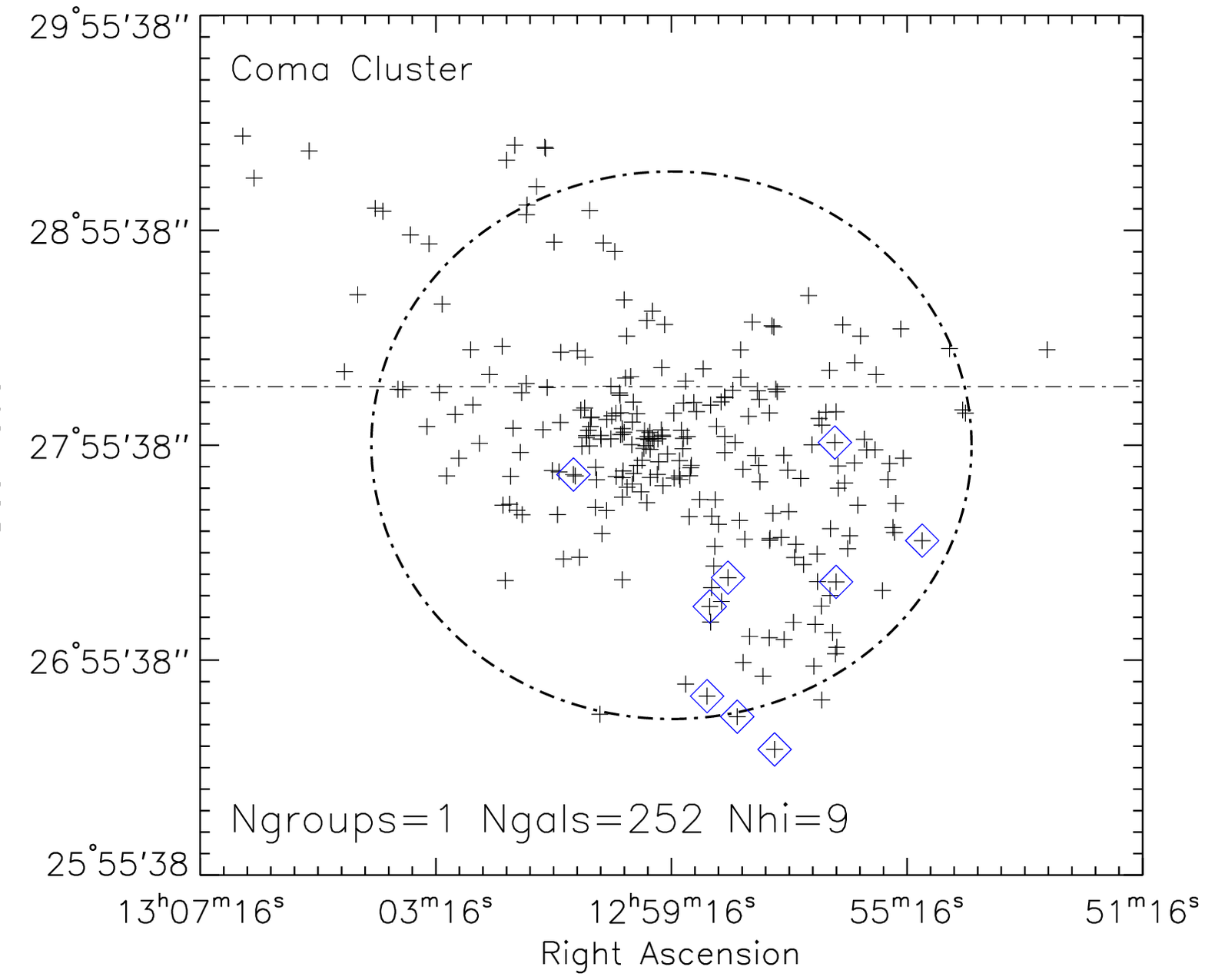}
\caption{Same as Figure \ref{pos67} but for the Coma Cluster.  In this case the circle corresponds a virial radius estimate of 2.3 Mpc \citep{Girardi98}.  The horizontal line corresponds to the declination limit of the $\alpha.40$ catalog.  Future ALFAFA data releases will cover the full extent of Coma Cluster. Despite the incompleteness, it is clear that the center of Coma is very \hi\ deficient and there are few \hi\ sources within half the virial radius.}
\label{poscoma}
\end{figure}

Figure \ref{poscoma} shows the distribution of galaxy members in Coma.  The dot-dashed horizontal line corresponds to the declination limit of the $\alpha.40$ coverage so \hi\ objects above this declination cannot be plotted.  Nonetheless, we include the Coma Cluster to demonstrate the extreme case: in the most massive dark matter halos, all \hi\ rich objects reside on the outskirts.  In this case, the circle corresponds to the virial radius of Coma (approximately 2.3 Mpc; \citealt{Girardi98}).

\begin{figure}
\centering
\includegraphics[scale=0.5]{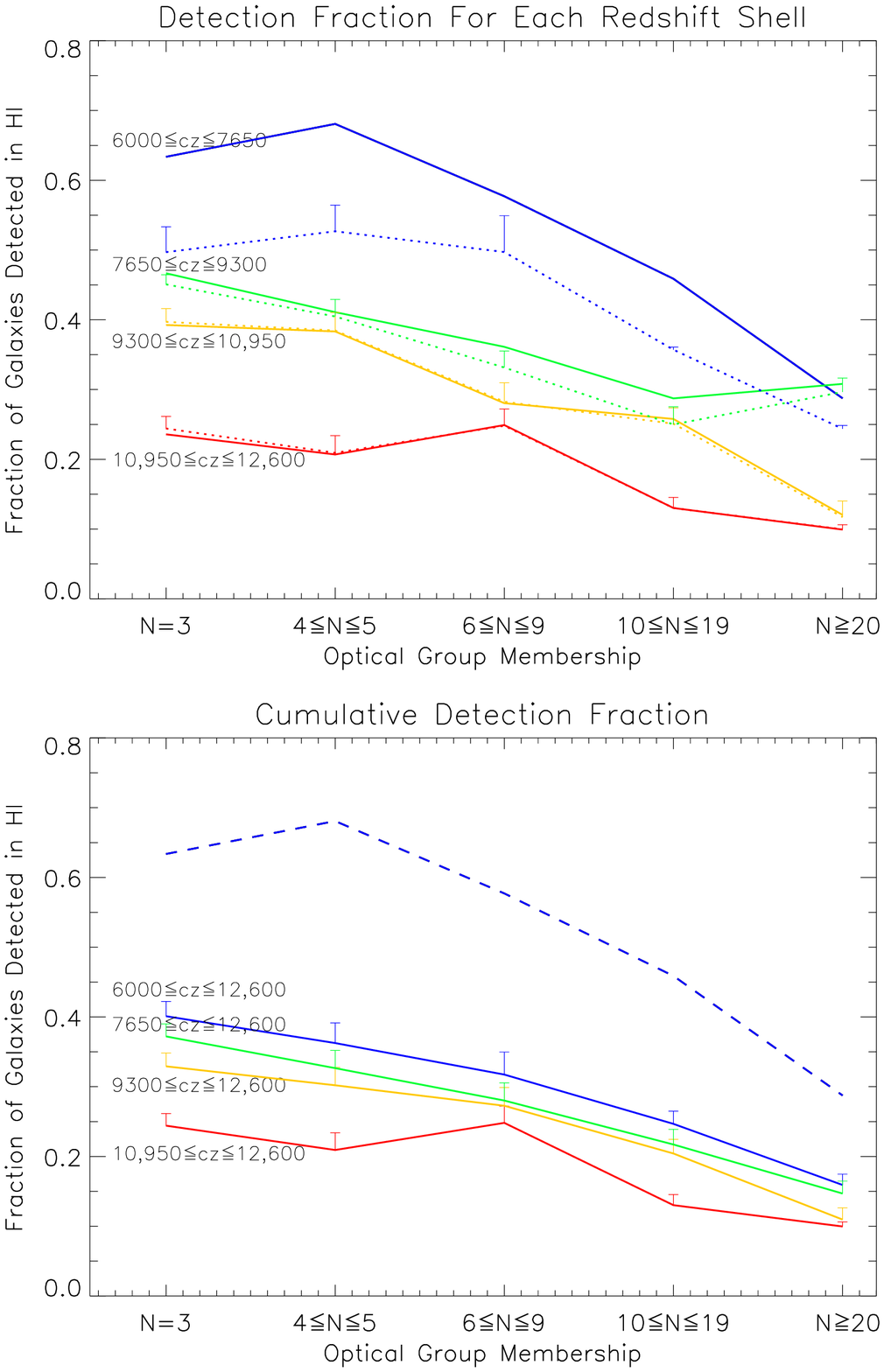}
\caption{Fraction of \hi\ detections as a function of group membership and plotted by redshift shell (blue, $6,­000-7,650$\kms; green, $7,650-­9,300$\kms; yellow, $9,300-­10,950$\kms; red $10,950­-12,600$\kms).  Top: the solid lines correspond to all \hi\ detected galaxies in a given shell.  The dotted lines correspond to a mass/volume-limited ALFALFA sample where $M\ge 10^{9.3}$\msun.  Bottom: The solid lines correspond to a mass/volume-limited ALFALFA sample, but include all galaxies from the lower redshift limit up to $12,600$\kms.  The dashed line is taken from the top panel for scale.  The vertical offset between solid lines of the same color suggest that we are affected by confusion and that the effect is greater at higher redshift where the resolution of the beam corresponds to a large project physical distance.  The vertical error bars in both panels demonstrate the correction if known ``blended'' sources (or similarly noted detections in Table 2 of \citealt{Haynes11}) are two sources within the same Arecibo beam.}
\label{hifrac}
\end{figure}

\subsection{The evolution of \hi\ in groups}
\label{evolution}

We recognize two major trends in how the gas content of galaxies varies with optical group membership.  First, the fraction of \hi\ detected galaxies decreases with increasing group membership.  Second, the spatial distribution of \hi\ detected galaxies changes with increasing group membership.  In groups with the fewest members ($N\leq9$), we find the \hi\ detected galaxies are concentrated towards the center or relatively evenly distributed through the group-sized volume.  As group membership increases, \hi\ detected members are less centrally concentrated, until the most massive groups are noticeably \hi\ deficient in their core, starting to resemble the \hi\ distribution of clusters (e.g.~Figure \ref{poscoma}).  Taken together, these two observations suggest that galaxies concentrated toward the center of massive groups have been members longer and have had more time to lose or use up their neutral atomic gas in the group environment.  Further, the spatial distribution suggests that the gas reservoir of groups is actively being replenished by fresh infall of gas rich galaxies from the surrounding environment--although we are unable to say whether this happens primarily through the accretion of individual galaxies, or small groups.

Additionally, we examine the group sample by redshift bin. The top panel of Figure \ref{hifrac} shows the fraction of group member galaxies that are detected in \hi\, as a function of optical group membership plotted for each redshift shell (color).  The solid lines correspond to the approximately flux-limited \hi\ sample.  We see that fewer galaxies are detected in \hi\ in more massive halos at all redshifts, but the fractional difference between low and high mass halos changes most dramatically in the lowest redshift shell (solid blue), while the highest redshift shell (solid red) has a relatively flatter slope.  The redshift range covered by the bins ($z=0.02-0.042$) is not enough to demonstrate cosmologically significant evolution.  However, because the optically defined groups come from a volume-limited sample, the only fundamental difference between groups in different redshift ranges is the \hi\ mass limit of the ALFALFA detections ($M_{HI} \gtrsim 10^{8.9} M_{\odot}$ at $cz_{\odot}=6,000-7,650$\kms\ to $M_{HI} \gtrsim 10^{9.3} M_{\odot}$ at $cz_{\odot}=10,950-12,600$).  The low \hi\ mass galaxies are already missing from the groups at $cz_{\odot}=10,950-12,600$ because ALFALFA is not sensitive to them at that distance.  Therefore, the fact that the slopes are flattening with increasing redshift implies that low \hi\ mass objects preferentially lose their gas first in the group environment. 

To test this we constructed a volume-limited \hi\ sample by applying the \hi\ mass sensitivity of the highest redshift shell to the full sample.  The dashed lines on Figure \ref{hifrac} correspond to the fraction of galaxies detected with $M_{HI} \gtrsim 10^{9.3} M_{\odot}$.  The difference between the full sample (solid) and the mass/volume-limited sample (dashed) is most dramatic in the lowest redshift shell (blue), which was sensitive to the lowest \hi\ mass objects.  However, the mass/volume-limited \hi\ sample still shows a trending vertical offset between the \hi\ detected fractions in different redshift shells.  The vertical error bars attempt to represent the uncertainty due to potentially confused sources, but additional sources of uncertainty include the choice of linking length to determine proximity members.

From the $\alpha.40$ catalog, only an estimated 5-9\% of \hi\ detections suffer from confusion.  In the bottom panel of Figure \ref{hifrac} we attempt to combat these small number statistics by plotting the fraction of \hi\ detections in cumulative redshift volumes.  The solid blue line represents the fraction of galaxies detected in \hi\ for the mass/volume-limited \hi\ sample over the full redshift range, $6000\ge cz\ge12,6000$.  The red solid line is only the highest redshift shell.  The intermediate lines are the cumulative detection fraction between the lower limit of the previously defined redshift shell, and the upper limit of our sample.  The goal of plotting the data in this way was to understand the vertical offset between the solid blue and red lines, since we have removed the dependence on \hi\ mass sensitivity.  The vertical error bars are again an attempt to represent the uncertainty based on confusion by assuming that each source noted in the ALFALFA catalog (Table 2; \citealt{Haynes11}) as being ``confused'', ``blended'', in a ``crowded field'', or a ``group'', is on average confused with one other source in the beam.  This assumption seems reasonable, since the vertical error bars of each successive cumulative sample are consistent with the next, more inclusive volume.  Confusion is more severe at higher redshift because the angular size of the beam corresponds to a greater physical distance, which accounts for the trend in vertical offset with redshift.

Based on the first panel of Figure \ref{hifrac}, we believe that the lowest \hi\ mass objects are losing their gas first in the group environment, however confusion contributes to fractionally fewer detected sources at high redshift.  In both panels of Figure \ref{hifrac} confusion is more common in small groups than in large groups.  Removing confusion strengthens our conclusions: the ability to resolve sources would increase the fraction of \hi\ detected sources preferentially in small groups and increase the number of low \hi\ mass objects in small groups by dividing the \hi\ mass among more sources.

\begin{figure*}
\centering
\includegraphics[scale=0.5]{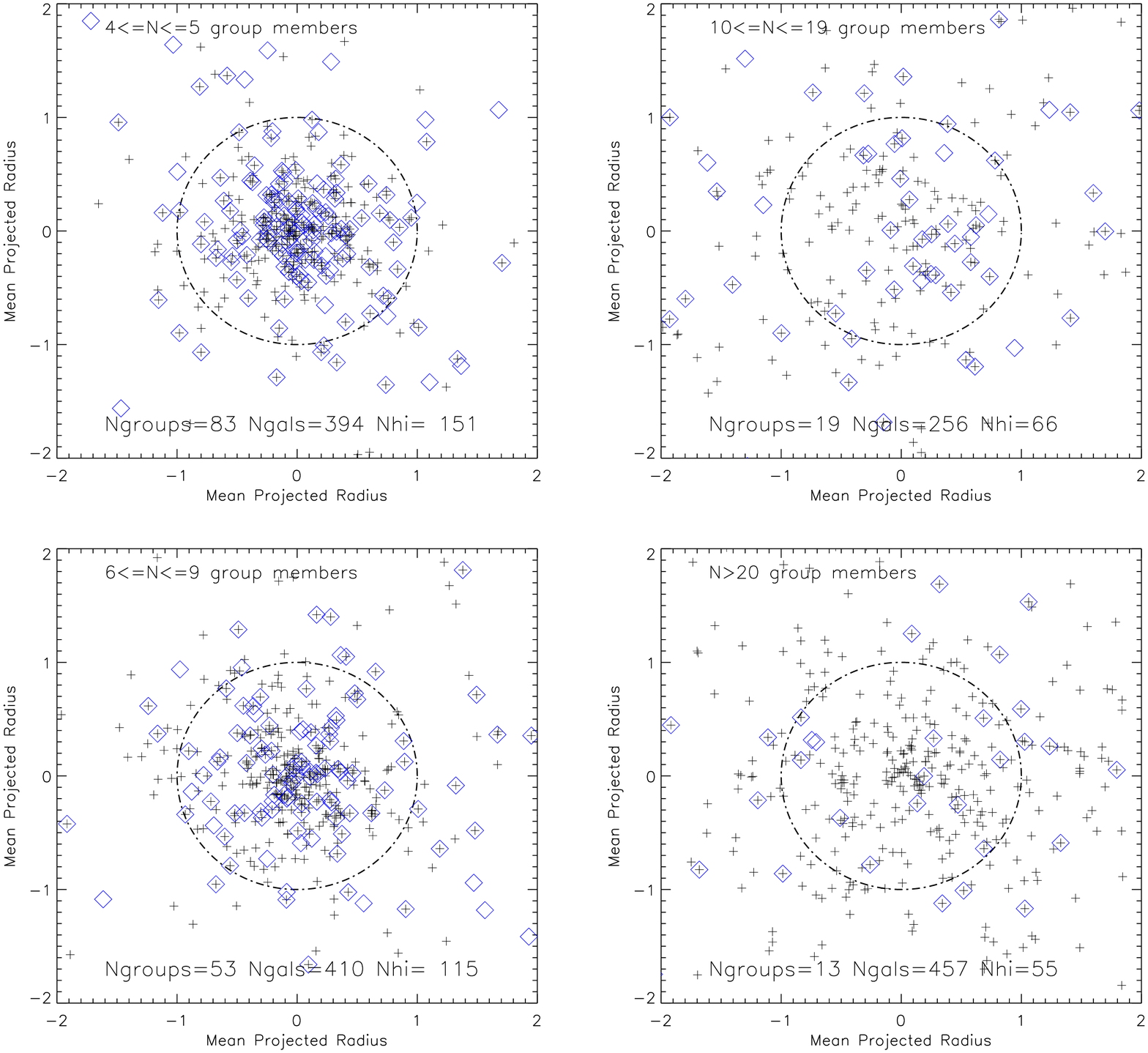}
\caption{The relative positions of group member galaxies are plotted for groups in the redshift shell $cz_{\odot}=9,300-10,950$\kms, but scaled to the mean projected radius of each group, respectively.  The symbols are the same as for Figures \ref{pos67} - \ref{pos1012}.  The dot-dashed circle corresponds to a mean projected radius normalized to 1.}
\label{rrms}
\end{figure*}

\begin{figure*}
\centering
\includegraphics[scale=0.5]{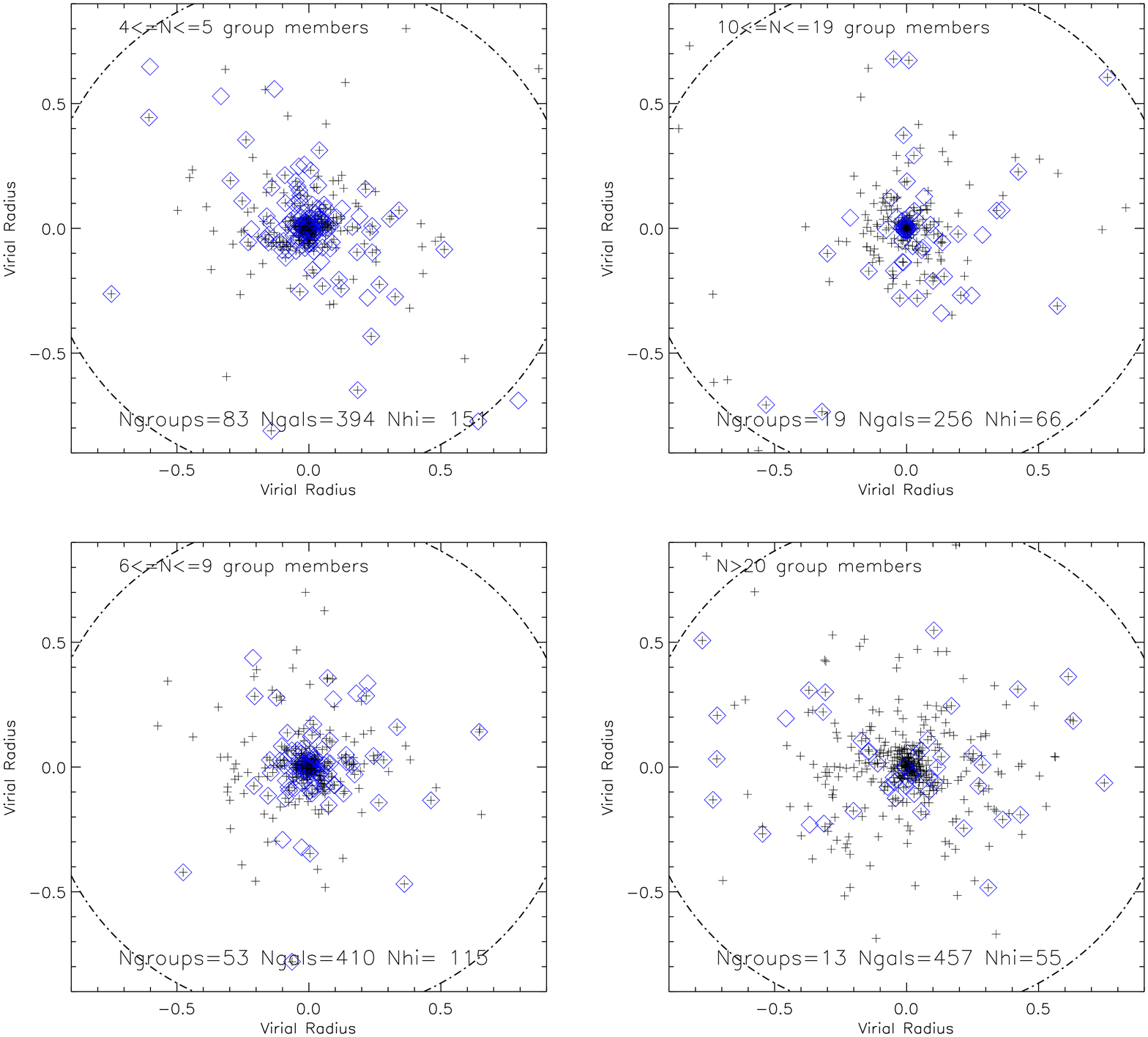}
\caption{The relative positions of group member galaxies are plotted for groups in the redshift shell $cz_{\odot}=9,300-10,950$\kms, but scaled to the virial radius of each group, respectively. The symbols are the same as for Figures \ref{pos67} - \ref{pos1012}.  The dot-dashed circle corresponds to a virial radius normalized to 1.  Only a very small fraction of group member galaxies reside outside the virial radius and have not been plotted.}
\label{rvir}
\end{figure*}

\subsection{\hi\ distribution scaled to group/virial radius}

The mean rms radius of the optical groups, as reported by \citet{Berlind09}, varies by a factor of 2.5 between $N=4-5$ and $N\geq20$ halos.  In order to remove any size dependence from the interpretation of our results and to compare groups of different membership in a consistent way, we scaled the groups by their projected rms radius and re-plotted the distribution of group member galaxies (Figure \ref{rrms}).  In this case, the dot-dashed circle corresponds to the normalized rms group radius.  We only show the results from one redshift shell but the same hold true for all redshift ranges.  Again, we find that the most massive groups are \hi\ deficient at their core.

Scaling by the virial radius is a more intuitive and physically motivated comparison between groups of different sizes.  However, estimating the virial radius of these groups is observationally difficult.  First, we cannot directly measure the underlying dark matter halo mass: groups with fewer than 10 members do not have enough points to estimate a velocity dispersion from spectroscopic observations, and we lack X-ray observations by which to measure the hot gas at the center of the group potential.  In fact, the majority of these groups are low mass and likely do not have an X-ray emitting intragroup medium.  Second, as evidenced by the \hi\ spatial distribution, infall is ongoing and therefore, groups are likely not virialized, causing us to overestimate the mass, or underestimate the virial radius.

Nonetheless, to estimate virial radius, we start with the virial theorem and assume a reasonable mass-to-light ratio to convert the Mr18 $r$-band group absolute magnitudes to total halo mass.  It has been found that in order to be virialized, groups must have $M/L=100-350 (M/L)_{\odot}$ and clusters have a mass-to-light ratio of order ($M_{vir}/L)_{cl}=350\pm70~h^{-1} (M/L)_{\odot}$ \citep{Carlberg96,Bahcall00,vandenBosch03}.  We have chosen a simple global value of $M/L=350$ to use for all group halos which we justify in Section \ref{hod}.  The value of $\sigma$ is taken as the line-of-sight velocity dispersion ($\sigma_v$) given by \citet{Berlind09} such that we evaluate the virial radius as:
\begin{equation}
M\sim\frac{3R\sigma^2}{G} \, \rightarrow \, R\sim\frac{350L_rG}{3\sigma_v^2}
\end{equation}
\citet{Weinmann06} also use luminosity as a proxy for mass and allege that this is a better estimate than conventional velocity dispersion \citep{Yang05}. 
Figure \ref{rvir} shows the groups plotted again, but scaled to the calculated virial radius (dot-dashed line).  

\subsection{Concentration of \hi\ and optical group members}
\label{concentration}

\begin{figure}
\centering
\includegraphics[scale=0.5]{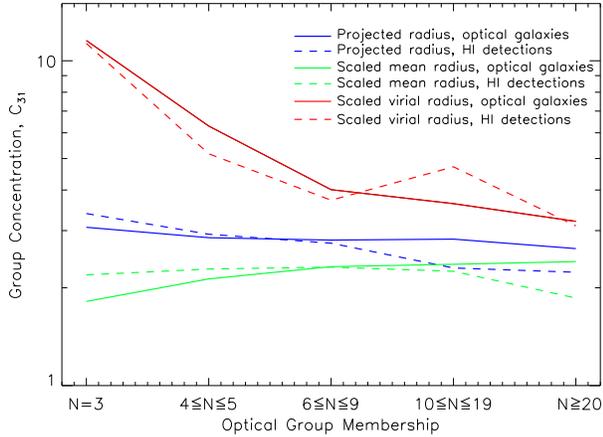}
\caption{The concentration of optical galaxies (solid lines) and \hi\ detected galaxies (dashed lines) in groups are plotted as a function of group membership.  The concentration is the ratio of the radius within which fall 75\% of group member galaxies versus 25\% of group member galaxies.  A larger value for $C_{31}$ corresponds to higher concentration. The relative concentration between optical and \hi\ detected group members switches around groups of $6\le N\le9$ members, which corresponds to a halo mass of $\sim10^13.1$\msun.}
\label{concen}
\end{figure}

In an attempt to quantify the relative concentration of groups we have used a modified version of $C_{31}$ which we define as the ratio between the radius within which reside 75\% of group members and the radius within which reside 25\% of the group members. We cannot do this for individual groups with membership much below 10, so we have used the all groups within an optical membership bin to effectively compute a mean concentration index.  In Figure \ref{concen}, we plot the modified $C_{31}$ as a function of group membership for the groups' projected radius (blue; e.g. Figures \ref{pos67}-\ref{pos1012}), the groups scaled to their mean geometric radius (green; Figure \ref{rrms}), and scaled to the virial radius (red; Figure \ref{rvir}) for the optical galaxies (solid lines) and the \hi\ detections (dashed lines).  

The projected group radius (blue) measures the spatial concentration of group members ``at face-value''.  Using this index, the optical galaxies are $\sim17$\% less concentrated in the largest mass halos compared to the smallest, while the \hi\ galaxies are $\sim52$\% less concentrated.  By comparison, when groups are normalized to the mean geometric radius (green), the trend is modified: groups which are spatially more extended relative to the mean are scaled down and become more concentrated while relatively smaller groups appear less concentrated.  Finally, the scaled virial radius index shows a strong decrease in concentration with increasing optical group membership.  The virial radius is strongly affected by the estimated velocity dispersion of the groups.  The majority of small groups have unrealistically small velocity dispersions (frequently of order tens of \kms) because their halos are not well sampled by only 3-9 members.  This results in an overestimate of the virial radius and an artificially high concentration calculated for the smallest groups.

Perhaps the most interesting result from an analysis of the group concentrations is that for both the projected group radius and scaled mean geometric radius indices, there is a transition around intermediate-sized groups where the \hi\ detections go from being more concentrated than the optical members in small group halos to less concentrated in large group halos.  These groups tend to have estimated halo masses around $10^{13.0-13.2} M_{\odot}$.  Interestingly, this is the halo mass at which \citet{McGee09} predict the environment starts to affect observed galaxy properties.  As will be shown in Section \ref{hod}, this is also the approximate halo mass at which the \hi\ selected halo occupation distribution starts to seriously deviate from a power law distribution.



\subsection{Halo Occupation Distribution}
\label{hod}

The halo occupation distribution (HOD) describes the mean number of galaxies per halo of a given mass.  In cosmological simulations, the HOD is derived in the context of the conditional luminosity function, and the results provide a test for our understanding of galaxy and structure formation, and galaxy clustering.  Much work has been done on this with regards to the stellar content of galaxies (color, mass, morphology, etc.;
).  Recent simulations have been extended to investigate the distribution of the cold gas component in dark matter halos as well \citep{Kim11}.  Here we provide the first observational comparison with simulations of the \hi\ selected HOD.

In simulations the difficulty is in populating dark matter halos with realistic galaxies.  Observationally, the difficulty is in using the galaxy distribution to estimate the mass and extent of dark matter halos.  We tested three methods to estimate the parent dark matter halo mass from the observed Mr18 group/cluster properties. First, we assumed a constant mass-to-light ratio for the individual galaxies consistent with what has been found for groups/clusters (of order $M_{vir}/L=100-350~h^{-1} (M/L)_{\odot}$).  Second, we assumed a varying mass-to-light ratio which is a function of halo mass (Equation 10 from \citealt{Yang05}). Neither of these approximations take directly into account the physical size or measured velocity dispersion of the groups, but use stellar luminosity as a proxy for halo mass.  Finally, we took the mean rms projected radius and the group velocity dispersion to calculate a ``dynamical mass'' using the virial theorem.  We tested each of these estimates by examining the distribution of masses which resulted for the Mr18 group/cluster catalog members.  We know that the most massive halo in this catalog is the Coma Cluster, and that the virial mass has been estimated from optical methods to be $\log(M/M_{\odot})\sim14.9$ \citep{Colless96,Girardi98}.  We expect the smallest halos to host groups to be $>10^{12} M_{\odot}$, with well-defined groups clustering around $10^{13.5} M_{\odot}$.

\begin{figure}
\includegraphics[scale=0.5]{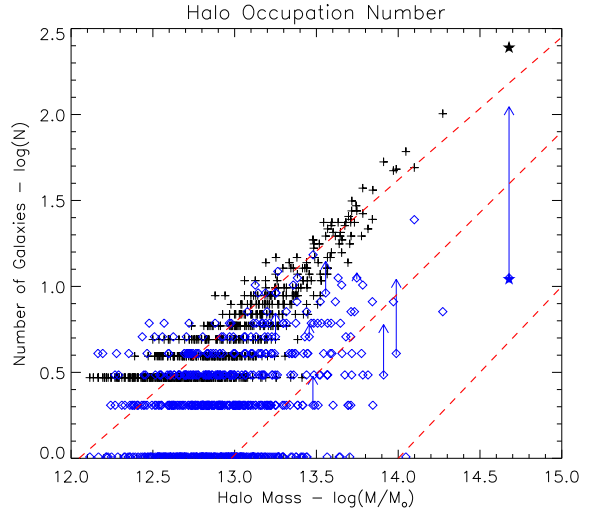}
\caption{The distribution of halo occupation numbers for each group.  Black crosses represent optical number counts for group members with $M_r<-18$.  Blue symbols represent \hi\ selected number counts for group members and include both PhotoObjID and proximity matches with $\log(M_{HI}/M_{\odot})\geq9.0$.  The symbols are offset by 0.02 in the vertical direction in order to better show the relative distributions.  The Coma Cluster is indicated by filled stars using the same color scheme for optical and \hi\ number counts.  Blue arrows are associated with groups that have at least one optical member outside the ALFALFA footprint.  The length of the arrow represents the halo occupation number if every optical galaxy outside the footprint was detected with $\log(M_{HI}/M_{\odot})\geq9.0$.  The red dashed lines represent satellite halo occupation distributions for \hi\ sources from three GALFORM models presented in \citet{Kim11}. In order from top to bottom they are: Font08, Bow06, MHIBow06.}
\label{hod+hi}
\end{figure}

Figure \ref{hod+hi} shows the optical and \hi\ halo occupation number for the group/cluster catalog calculated using the fixed mass-to-light ratio of $M/L=350~(M/L)_{\odot}$.  Black crosses represent the number of optical galaxies with $M_r < -18$ in each group.  Blue symbols represent the number of \hi\ detected galaxies with $\log(M_{HI}/M_{\odot})>9.0$ in each group.  Blue arrows are associated with groups that have one or more optical member outside the ALFALFA survey boundaries.  The length of the arrows represents the halo occupation number if every optical galaxy outside the footprint was detected with $\log(M_{HI}/M_{\odot})\geq9.0$.  Although the fixed mass-to-light ratio slightly underestimates the mass of Coma, the low and intermediate mass sized halos fall in the expected range of values.  By comparison, using the varying mass-to-light ratio of \citet{Yang05}, the lowest mass groups are underestimated at $10^{11} M_{\odot}$ and the mass of Coma is overestimated by an order of magnitude, although the intermediate mass groups agree well with the expected values.  The discrepancy at the high mass end is likely due to the fact that the pairwise peculiar velocity and group multiplicity analysis of \citet{Yang04} predict $(M/L)_{cl}=900~h^{-1} (M/L)_{\odot}$--significantly higher than shown by observations \citep{Carlberg96,Bahcall00}.  An explanation may lie in Figure 3 of \citet{Berlind06} which demonstrates that one cannot recover realistic groups while simultaneously optimizing for an unbiased velocity dispersion, size, and multiplicity function.  Meanwhile, using the rms projected radius and velocity dispersion greatly underestimates the virial mass for all halos.

Fitting an HOD curve to these values for direct comparison to simulations is a non-trivial problem.  Aside from the uncertainty of our virial mass estimate, there are a number of factors which are either not included in this plot, or which cannot be measured observationally but which are known in simulations.  In particular, we cannot include halos with pairs or single galaxies.  It also excludes halos with multiple galaxies that fall below the $r$-band magnitude limit--not even counting them as halos with zero galaxies meeting the optical criteria.  Comparison between our results and simulations are limited to the high mass end of the HOD.  At the very low mass end, not probed by this study, the HOD is dominated by the ``central'' galaxy population and halos containing only one massive galaxy.  At the intermediate and high mass end, the HOD is dominated by the ``satellite'' galaxy population, and we expect better agreement.

In Figure \ref{hod+hi}, we compare our results with the HOD prediction for \hi\ selected galaxies with $M\ge10^9 M_{\odot}$ of three GALFORM cosmological models from \citet{Kim11} (red dashed lines). From top to bottom, these are known as Font08, Bow06, MHIBow06.  Bow06 was designed to match the optical luminosity function of galaxies.  Font08 is meant to have a better treatment of hot gas stripping in dark matter halos.  The resulting simulation has more faint blue galaxies and better color agreement with optical observations, but both Bow06 and Font08 models over-predict the HIPASS \hi\ mass function (HIMF; \citealt{Zwaan05}).  MHIBow06 was adjusted to match the HIPASS HIMF, but well under-predicts the data here.  This particular discrepancy may arise from the fact that the ALFALFA extrapolated Schechter function fit to the HIMF predicts an order of magnitude more galaxies at the high mass end \citep{Martin10}.  Our results appear to indirectly support that prediction since our HOD estimates were generated by selecting galaxies with $M\ge10^9 M_{\odot}$.  

So far, the best fit cosmological model lies somewhere between Bow06 and Font08.  However, Figure \ref{hod+hi} may suggest that the occupation number of the \hi\ selected population does not rise as steeply with halo mass as suggested by any of the models, and may even start to flatten at intermediate mass halos ($\ge10^{13}$\msun).  Our interpretation is cautious one, but it is consistent with the conclusions of \citep{McGee09} that dominant environmental processes begin in halos of about this mass.  Further, it suggests that simulations do not yet properly treat the processing of gas in the group environment.

In some of our Mr18 groups, the number of group members is greater when selected by \hi\ than by $r$-band optical magnitudes due to the proximity matches.  We decided to not change the group membership statistics in our previous analysis to reflect this in order to remain consistent in our method of estimating halo masses.  We have plotted the halo occupation numbers in ways that the HOD have been consistently expressed from simulations: selecting galaxies brighter than a certain optical magnitude, or above an \hi\ mass cutoff.  The differences between the optical and \hi\ halo occupation numbers not only shows that gas is being processed in galaxies in groups, but also shows that the two approaches select qualitatively different galaxies at the low mass end depending on whether they are selected on stellar or gas content.

\section{Discussion}
\label{discussion}

Figures \ref{pos67}-\ref{pos1012} show the most convincing evidence to date that not only is \hi\ gas being pre-processed in galaxy groups, but reveal clues to how it is happening in a spatially resolved way.  Prior evidence for pre-processing of the \emph{gas content} in the group environment comes from estimates of the \hi\ mass function (HIMF) which historically have been derived from individual groups or small group samples (e.g.~\citealt{Verheijen00,deBlok02,Kovac05,Freeland09,Kilborn09,Pisano11,Hess11}).  Despite the small number statistics, calculations of the group HIMF reliably suggest a flatter slope at the low mass end than the field or global HIMF.  If only 25\% of ALFALFA detections are in groups/clusters, then the global HIMF is dominated by \hi\ detections which reside in the field, and groups are deficient in low \hi\ mass galaxies compared to the field.  In further support of this, we have demonstrated that low \hi\ mass objects lose their gas first in the group environment, consistent with a group HIMF that has a flatter low mass slope than the field \citep{Freeland09,Kilborn09,Pisano11,Hess11}, and with the idea that dwarf galaxies can be stripped of their \hi\ gas in the group environment (e.g.~\citealt{Marcolini03,Hester06,Freeland11}).  In contrast, \citet{Zwaan05} concluded that the HIPASS HIMF slope becomes steeper with increasing density, however, the density metric used was the distance to the fifth nearest \hi\ detected neighbor.  Our results show that \hi\ detected sources are not well clustered on small scales and that most \hi\ detections correspond to poor groups, with rich groups having a lower fraction of \hi\ detections.

Precisely how galaxies lose their gas in unclear.  The similarity between increasingly large groups and clusters, in which \hi\ rich galaxies are found to reside on the outskirts, emphasizes the continuum of dark matter halo masses that exist between groups and clusters.  It also gives cause to re-evaluate physical processes which have been previously reserved for transforming galaxies in clusters, namely ram pressure stripping, may also be important in the group environment.  However, it is unclear whether the conditions of the intragroup medium is sufficient for ram pressure stripping.  Traditional wisdom is that strong gravitational encounters and mergers are more common in groups than in clusters and that these interactions are the dominant mechanism contributing to galaxy evolution.  Ram pressure stripping can remove the hot gas halo of galaxies in groups (starvation), but similar trends in the cold gas may also be explained by viscous stripping \citep{Rasmussen12}.  Tantalizingly, \citet{Freeland11} have shown from studies of bent radio jets in galaxy groups that the intragroup medium density can be greater than previous thought ($0.2-3\times10^{-3}$ cm$^{-3}$ at $15-700$ kpc from the group center).  This is sufficient to completely strip the gas from dwarf galaxies, and begins to explain why low \hi\ mass systems are less common in massive groups.

Nurture is clearly at work in the group environment.  We observe that fewer HI sources are detected in larger groups than in smaller ones.  This fact by itself does not support either nature or nurture over the other.  Assuming large groups are older and therefore have had more time to assemble, if galaxies form in-situ they have also had more time to form stars, and longer to exhaust their gas.  However, the presence of \hi\ may be a good tracer of substructure \citep{BravoAlfaro00}, and when we examine the most massive groups on an individual basis they frequently appear lopsided in their \hi\ distribution.  We propose a scenario in which, early in life, groups are small and gas rich.  Groups and their parent dark matter halos grow through the accretion of individual gas rich galaxies from the field, or through merging with smaller groups.  As galaxies experience the group environment, their gas is ``processed''.  As the group continues to grow, the stellar content of galaxies remains.  Meanwhile, galaxies that have been in the group the longest have lost their gas or turned it into stars, and the cold \hi\ gas content of increasingly large groups is dominated by the \hi\ content brought in by new arrivals.  These may be individual galaxies, or smaller groups.  If we assume that groups are growing from $N=3$ to $N \ge 20$, then this says that low HI mass galaxies preferentially process their gas first possibly through interactions, starvation, ram pressure stripping, or star formation.  The relative importance of these processes may vary with different halo masses, but the fate of the cold neutral gas may be traceable--through measurements of star formation activity in individual galaxies, or the accumulation of warm/hot diffuse UV/X-ray emitting gas in the intragroup medium \citep{Rasmussen12,Desjardins13}.

\section{Conclusions}
\label{conclusions}

We have used publicly available data from complementary SDSS DR7 Mr18 group/cluster and ALFALFA $\alpha.40$ catalogs to study the cold neutral gas content of groups, clusters, and their constituent galaxies.  The Mr18 catalog is a magnitude/volume limited catalog, complete for groups with $N\geq3$ over the range $z=0.02-0.042$.  It is optimized to fit the size and multiplicity function for groups of 10 or more members \citep{Berlind06,Berlind09}.  ALFALFA $\alpha.40$ is an approximately flux-limited catalog which detects \hi\ rich objects over the same volume \citep{Haynes11}.  The stellar content best traces the underlying parent dark matter halos, thus we have used the optically determined groups to identify large scale structures.  Meanwhile, we use the radio catalog to trace how the cold gas content of galaxies has been modified by their environment.  By investigating both flux-limited and mass/volume-limited \hi\ samples, we have demonstrated that confusion is a small, but noticeable effect in our analysis, but works to enhance our conclusions.

For the first time, we have demonstrated evolution in the spatial distribution of the \hi\ content of galaxy groups as a function of their group membership, which we use as a proxy for the underlying dark matter halo mass.  Further, we study how the fraction of \hi\ detections changes in different redshift shells, which is effectively the same as selecting a different limiting \hi\ mass.  Our results can be summarized as follows:
\begin{enumerate}
\item{Only 24.8\% of \hi\ all detected sources reside in groups or clusters, compared to approximately half of all optically selected galaxies.}
\item{The spatial distribution of \hi\ rich galaxies evolves with group dark matter halo mass: \hi\ galaxies increasingly reside on the outskirts of groups as the optical membership increases.}
\item{The fraction of \hi\ detections decreases with increasing halo mass: galaxies at the center of groups have used or lost their gas through star formation, galaxy-galaxy interactions, starvation, or ram pressure stripping.}
\item{Low \hi\ mass galaxies preferentially lose their gas first in increasingly massive group halos.}
\item{The infall of gas rich galaxies replenishes the gas content of intermediate and high mass group halos.}
\end{enumerate}

Further, we attempt to present our results for the \hi\ content of groups in a context that can be compared to the halo occupation distribution (HOD) statistic derived from simulations, although we are limited to the intermediate and high mass end of halo masses, which are dominated by the satellite galaxy population. Our results indicate that the transition between gas rich and gas poor is happening in the group environment around a halo mass of $10^{13.0-13.2}$\msun, and the fresh infall of galaxies is important to the build-up of large scale structure and replenishing gas in these structures at the present epoch.

It is unclear whether the processing or loss of cold gas in groups is due to star formation, tidal interactions, starvation, viscous stripping, or ram pressure stripping.  Differentiating between these physical mechanisms requires deep, high resolution \hi\ observations to search for intragroup \hi\ clouds, tidal debris, swept back \hi\ disks, and H$\alpha$ or UV observations to spatially resolve and measure star formation.  As such, we have undertaken a campaign to study one such intermediate mass group known to exhibit segregation in the \hi\ gas content of its members and substructure.  Our results will be the subject of a future publication.

\section{Acknowledgments}

KMH would like to acknowledge the intangible contributions and support of CBH, JBH, SLH, JMS, CMW, NMG, BCE, AMH, KJO, and MAH throughout this research.  The authors would like to thank J.H. van Gorkom for useful discussions and the anonymous referee for remarks which improved the quality of this manuscript.

This research has been supported in part by the South African Research Chairs Initiative (SARChI) of the Department of Science and Technology (DST), the Square Kilometre Array South Africa (SKA SA), and the National Research Foundation (NRF). This work was partially supported by the National Science Foundation award AST-0730052 and support from the University of Wisconsin-Madison Graduate School.

The authors acknowledge the work of the entire ALFALFA collaboration team in observing, flagging, and extracting the catalog of galaxies used in this work.  The Arecibo Observatory is operated by SRI International under a cooperative agreement with the National Sci- ence Foundation (AST-1100968), and in alliance with Ana G. Mndez-Universidad Metropolitana, and the Universities Space Research Association.

Funding for the SDSS and SDSS-II has been provided by the Alfred P. Sloan Foundation, the Participating Institutions, the National Science Foundation, the U.S. Department of Energy, the National Aeronautics and Space Administration, the Japanese Monbukagakusho, the Max Planck Society, and the Higher Education Funding Council for England. The SDSS Web Site is http://www.sdss.org/.  The SDSS is managed by the Astrophysical Research Consortium for the Participating Institutions. The Participating Institutions are the American Museum of Natural History, Astrophysical Institute Potsdam, University of Basel, University of Cambridge, Case Western Reserve University, University of Chicago, Drexel University, Fermilab, the Institute for Advanced Study, the Japan Participation Group, Johns Hopkins University, the Joint Institute for Nuclear Astrophysics, the Kavli Institute for Particle Astrophysics and Cosmology, the Korean Scientist Group, the Chinese Academy of Sciences (LAMOST), Los Alamos National Laboratory, the Max-Planck-Institute for Astronomy (MPIA), the Max-Planck-Institute for Astrophysics (MPA), New Mexico State University, Ohio State University, University of Pittsburgh, University of Portsmouth, Princeton University, the United States Naval Observatory, and the University of Washington.

\end{document}